\documentclass[twocolumn,prl,preprintnumbers,tightenlines,nofootinbib,superscriptaddress]{revtex4}

\pdfoutput=1
\usepackage{amsmath,amssymb,amsfonts}
\usepackage{graphicx}
\usepackage{booktabs}
\usepackage{hyperref}
\usepackage{color}
\usepackage[sort&compress]{natbib}
\usepackage{tikz}
\usepackage{cleveref}
\newcounter{qnumber}

\newcommand{\drawsquare}[2]{\hbox{%
\rule{#2pt}{#1pt}\hskip-#2pt
\rule{#1pt}{#2pt}\hskip-#1pt
\rule[#1pt]{#1pt}{#2pt}}\rule[#1pt]{#2pt}{#2pt}\hskip-#2pt
\rule{#2pt}{#1pt}}

\newcommand{\Yfund}{\raisebox{-.5pt}{\drawsquare{6.5}{0.4}}}
\newcommand{\Yasymm}{\raisebox{-3.5pt}{\drawsquare{6.5}{0.4}}\hskip-6.9pt%
        \raisebox{3pt}{\drawsquare{6.5}{0.4}}}

\begin{document}

\title{Some Exact Results in Chiral Gauge Theories}

\author{Csaba Cs\'aki}
\email{ccsaki@gmail.com}
\affiliation{Department of Physics, LEPP, Cornell University, Ithaca, NY 14853, USA}

\author{Hitoshi Murayama}
\email{hitoshi@berkeley.edu, hitoshi.murayama@ipmu.jp, Hamamatsu Professor}
\affiliation{Department of Physics, University of California, Berkeley, CA 94720, USA}
\affiliation{Kavli Institute for the Physics and Mathematics of the
  Universe (WPI), University of Tokyo,
  Kashiwa 277-8583, Japan}
\affiliation{Ernest Orlando Lawrence Berkeley National Laboratory, Berkeley, CA 94720, USA}

\author{Ofri Telem}
\email{t10ofrit@gmail.com}
\affiliation{Department of Physics, University of California, Berkeley, CA 94720, USA}
\affiliation{Ernest Orlando Lawrence Berkeley National Laboratory, Berkeley, CA 94720, USA}

\begin{abstract} 
We analyze dynamics of chiral gauge theories based on the $SU(N)$ gauge group with one anti-symmetric tensor $A$ and $(N-4)$ anti-fundamentals $F_{i}$ when $N$ is odd. Based on the continuity to the supersymmetric gauge theories with anomaly-mediated supersymmetry breaking, we claim that the global $SU(N-4)$ symmetry is spontaneously broken to $Sp(N-5)$. There are $N-5$ massless fermions as a fundamental representation of $Sp(N-5)$, and another massless fermion, together saturating the anomaly matching conditions. When $N$ is even, the unbroken flavor symmetry is $Sp(N-4)$ while there are no massless fermions. Our result is different from the dynamics suggested by tumbling where the full $SU(N-4)$ symmetry is unbroken, but the tumbling picture can be modified via the addition of a second condensate to produce the symmetry breaking pattern predicted from our method.
\end{abstract}

\maketitle

\section{Introduction}

Non-abelian gauge theories \cite{Yang:1954ek} are the basis of our modern understanding of microscopic physics. Quantum Chromo-Dynamics (QCD) based on the $SU(3)$ gauge group is a prime example with vector-like particle content. In general, $SU(N)$ gauge theories with $N_{f}$ quarks in the fundamental representation are called QCD-like theories. Inspired by the light pions and Bardeen--Cooper--Schrieffer (BCS) theory of superconductivity \cite{BCS}, Nambu and Jona-Lasinio conjectured dynamical chiral symmetry breaking \cite{Nambu:1961tp,Nambu:1961fr}, which is now believed to be the correct dynamics of QCD-like theories.

However, the dynamics of chiral gauge theories are difficult to understand. There has been theoretical progress in representing chiral gauge theories on the lattice \cite{Ginsparg:1981bj,Kaplan:1992bt,Shamir:1993zy,Narayanan:1994gw,Luscher:1998pqa,Golterman:2000hr,Golterman:2004qv,Grabowska:2015qpk,Grabowska:2016bis}, which is a potential avenue for future numerical simulations. Yet they are numerically expensive and progress is slow. Arguably understanding the dynamics of chiral gauge theories is one of the most important open questions in quantum field theories.

While there is no established systematic approach there does exist a conjectured framework for the dynamics of chiral gauge theories called tumbling  \cite{Raby:1979my}. It postulates certain fermion bilinear condensates that dynamically breaks the gauge symmetry until the remaining gauge group becomes QCD-like. For example, an $SU(N)$ gauge theory with an anti-symmetric tensor $A$ and $(N-4)$ anti-fundamentals $\bar{F}_{i}$ was argued~\cite{Dimopoulos:1980hn} to break the gauge symmetry by the condensate $\langle A^{ab} \bar{F}_{i}^{b} \rangle = v^{3} \delta^{a}_{i}\neq 0$ to an $SU(4)$ gauge theory that confines. It assumes massless symmetric tensor composite fermions $A \bar{F}_{\{i,} \bar{F}_{j\}}$.  Even though the conjecture satisfies non-trivial 't Hooft anomaly matching conditions, it has never been clear if it is the correct understanding. 
Recently, the above proposals were examined  in \cite{Bolognesi:2020mpe,Bolognesi:2021yni}, via new discrete anomaly matching conditions \cite{Csaki:1997aw}, linked to the center symmetry of the gauge group \cite{Gaiotto:2017yup,Tanizaki:2018wtg,Bolognesi:2019fej}. We do not elaborate further on these consistency conditions, since they always hold for our proposal of the IR dynamics. This is because of its continuous connection to the supersymmetric theory.

In~\cite{Murayama:2021xfj}, a novel approach was proposed to study the dynamics of non-supersymmetric gauge theories via anomaly-mediated supersymmetry breaking (AMSB) \cite{Randall:1998uk,Giudice:1998xp}. It is based on the Weyl compensator field
\begin{align}
	\Phi = 1 + \theta^{2} m,
	\label{eq:compensator}
\end{align}
where $m$ dictates the size of the supersymmetry (SUSY) breaking. The UV theory has mass for squarks and gauginos, which decouple from dynamics when $m$ is increased, and it is therefore continuously connected to non-supersymmetric gauge theories. Due to the ultraviolet-insensitivity of the AMSB \cite{Pomarol:1999ie,Boyda:2001nh}, the dynamics can be studied using the particle content and interactions at each energy scale. In particular, a consistent picture was obtained for QCD-like theories \cite{Murayama:2021xfj}. For other approaches to extrapolating from supersymmetric theories to their non-SUSY counterparts, see, for example, \cite{Aharony:1995zh},\cite{Cheng:1998xg}. 

In this paper, we apply the AMSB methodology to analyze the dynamics of non-supersymmetric chiral gauge theories. We begin the discussion with the simplest and most well-known chiral gauge theory: $SU(5)$ with an anti-symmetric tensor and an anti-fundamental Weyl fermion. Its supersymmetric version is well-known to break SUSY dynamically, though the actual dynamics is not calculable. We point out that in the SUSY breaking minimum we expect a massless composite fermion, which is expected to persist in the non-supersymmetric theory after adding AMSB. Next we analyze the general $SU(N)$ ($N=2n+1$ odd) theories with an anti-symmetric tensor and $N-4$ anti-fundamentals. Again the SUSY version with AMSB can be worked out exactly, leading to the dynamical breaking of the $SU(N-4)\times U(1)$ global symmetry to $Sp(N-5)\times U(1)$, as well as massless fermions in the fundamental and singlet representations of $Sp(N-5)$. This picture continuously connects to the non-SUSY limit, while it does not agree with the simplest tumbling predictions. We show however that one can extend the tumbling picture by adding another condensate in the second most attractive channel to obtain a symmetry breaking pattern that agrees with the non-SUSY limit of the AMSB approach. Finally we also discuss the case of even $N=2n$ with no massless fermions content.

\section{Anomaly Mediation}

Anomaly mediation of supersymmetry breaking (AMSB) is parameterized by a single number $m$ that explicitly breaks supersymmetry in two different ways. One is the tree-level contribution based on the superpotential
\begin{align}
	{\cal L}_{\rm tree} &= m \left( \phi_{i} \frac{\partial W}{\partial \phi_{i}} - 3 W \right)
	+ c.c. \label{eq:AMSBW}
\end{align}
The other is the loop-level supersymmetry breaking effects in tri-linear couplings, scalar masses, and gaugino masses,
\begin{align}
	A_{ijk} (\mu) &= - \frac{1}{2} (\gamma_{i} + \gamma_{j} + \gamma_{k})(\mu) m, \\
	m_{i}^{2}(\mu) &= - \frac{1}{4} \dot{\gamma}_{i}(\mu) m^{2}, \\
	m_{\lambda}(\mu) &= - \frac{\beta(g^{2})}{2g^{2}}(\mu) m.
\end{align}
Here, $\gamma_{i} = \mu\frac{d}{d\mu} \ln Z_{i}(\mu)$, $\dot{\gamma} = \mu \frac{d}{d\mu} \gamma_{i}$, and $\beta(g^{2}) = \mu \frac{d}{d\mu} g^{2}$. When the gauge theory is asymptotically free, $m_i^2>0$ which stabilizes the theory against run-away behaviors.
Note that Eq.~\eqref{eq:compensator} also breaks the $U(1)_R$ symmetry explicitly and hence we do not need to study its anomaly matching conditions.

\section{$SU(5)$ with $A({\bf 10})$ and $\bar{F}({\bf \bar{5}})$}

This is the simplest and most well-known chiral SUSY gauge theory which has a non-anomalous $U(1)_{5}$ symmetry $A(+1)$, $\bar{F}(-3)$, see Table~\ref{tab:SU(5)}. This theory breaks supersymmetry dynamically \cite{Affleck:1983vc}. (See \cite{Leigh:1997sj} for how this theory is related to the 4-1 and 3-2 models.)
Even though the theory is intrinsically strongly coupled, it was shown that adding a pair of $F+\bar{F}$ allows for a weakly-coupled analysis \cite{Murayama:1995ng} (see also \cite{Poppitz:1995fh,Pouliot:1995me}).  Once a mass term is added to the pair,  one can explicitly calculate the actual SUSY breaking minimum of the potential. See the Appendix for details of the analysis.

Surprisingly one of the most important features of the SUSY breaking minimum has so far (to the best of our knowledge) not been discussed in the literature. We are finding that in spite of the strong dynamics the $U(1)_{5}$ global symmetry remains unbroken, implying the presence of a massless fermion that can be identified with the composite fermion  $A \bar{F} \bar{F}$ \footnote{More explicitly, the massless fermion is given by $(\epsilon^{\alpha\beta}A^{ab}_\alpha\bar{F}_{2a\beta})\bar{F}_{2b\gamma}$, where $\alpha,\beta,\gamma=1,2$ are spinor indices.}. This massless fermion exists in addition to the massless Goldstino required for SUSY breaking. It has charge $-5$ and one can readily check that it saturates all 't Hooft anomalies:
\begin{align}
U(1)_{5}\,{\rm gravity}^{2}:&~~~~~~{\bf 10}~(+1)+{\bf \bar{5}}~(-3)~= ~(-5)~,\nonumber\\[5pt]
U(1)_{5}^{3}:&~~~~{\bf 10}~(+1)^{3}+{\bf \bar{5}}~(-3)^{3}\, = ~(-5)^{3} .
\label{eq:U(1)}
\end{align}
This non-trivial 't Hooft anomaly matching is already a very strong argument in favor of the existence of the unbroken $U(1)_{5}$ symmetry, which as explained above can be independently verified by analyizing the theory with the extra flavor added, which is done explicitly in the appendix. 

Since supersymmetry is already spontanneously broken, adding small explicit supersymmetry breaking via AMSB does not change the dynamics, while the goldstino acquires a mass. On the other hand, the other massless fermion $A \bar{F} \bar{F}$ is protected because of the anomaly matching condition and remains massless even in the non-supersymmetric theory. The appearance of the unbroken $U(1)_{5}$ and the corresponding massless fermion is our first example of exact results in chiral gauge theories. Extrapolating $m \gg \Lambda$, we expect the same massless fermion remaining in the theory while the goldstino decouples.

It is curious that our analysis can be nicely reconciled with results suggested by tumbling \cite{Georgi:1979md,Raby:1979my}. In that approach one would postulate that the operator $\epsilon_{abcde}A^{bc} A^{de}$ corresponding to the Most Attractive Channel (MAC) condensate \footnote{When discussing the tumbling approach, we will use the heuristic notion of gauge-dependent condensates $\left<\Phi_a\right>$, which is formally at odds with Elitzur's theorem \cite{Elitzur:1975im}. These should be read as stand-in for the theory being in a Higgs phase.}, breaking the $SU(5)$ gauge symmetry to $SU(4)$, while retaining a diagonal global $U(1)$. Based on our later discussions, we suggest that a less attractive channel $A^{ab} \bar{F}_b$ also condenses and contributes to the same gauge symmetry breaking from $SU(5)$ to $SU(4)$. Under the unbroken $SU(4)_{\text{gauge}}\times U(1)_{\text{global}}$, $A$ decomposes as ${\bf 6}_0+{\bf 4}_{-5/2}$ while $\bar{F}$ as ${\bf \bar{4}}_{5/2} + {\bf 1}_{-5}$. The degrees of freedom charged under $SU(4)$ are vector-like, and so they condense just like in QCD and become massive. However, $\bf 1_{-5}$ is chiral, and remains as a massless fermion. It can be identified as $A \bar{F} \bar{F}$ with our assumption of the $A^{ab} \bar{F}_b$ condensate. Note that for general $SU(N)$ our method leads to predictions that differ somewhat from those in~\cite{Raby:1979my}. However we will see that the tumbling scheme can be augmented to produce results consistent with ours.

\begin{table}[t]
	\centerline{
	\begin{tabular}{|c|c|c|} \hline
	& $SU(5)$ & $U(1)_{5}$ \\ \hline
	$A$ & $\Yasymm$ & $+1$ \\ \hline
	$\bar{F}$ & $\overline{\Yfund}$ & $-3$ \\ \hline \hline
	$A \bar{F} \bar{F}$ & ${\bf 1}$ & $-5$ \\ \hline
	\end{tabular}
	}
	\caption{Particle content of the $SU(5)$ theory with one anti-symmetric $A$ and one anti-fundamental $\bar{F}$. The composite fermion $A \bar{F} \bar{F}$ matches all anomalies.}\label{tab:SU(5)}
\end{table}

\section{$SU(N)$ with $A$ and $(N-4)$ $\bar{F}$}

Next we consider the generalization of the above $SU(5)$ theory to $SU(N)$ with $N=2n+1$ odd. Table~\ref{tab:SU(N)} shows the matter content and the symmetries of the theory. In the SUSY limit, the theory has the $D$-flat direction 
\begin{align}
	\lefteqn{
	 A A^{\dagger} + A^{\dagger} A - \bar{F}_{i}^* \bar{F}_{i}^{T} =0, 
	 } \\
    A = \frac{\varphi}{\sqrt{2}} & \left( \begin{array}{c|c}
        J_{(N-5)} & 0 \\ \hline
        0 & 0_{5\times 5}
        \end{array} \right)\ , 
	&\bar{F} = \varphi \left( \begin{array}{c|c}
		I_{(N-5)} & 0  \\ \hline
		0 &0_{5\times 1} 
		\end{array} \right)\ ,
\end{align}
where $J_{(N-5)}=i\sigma_2 \otimes I_{(N-5)/2}$. As we will see below, in our scenario the flat direction will ultimately be stabilized at large VEVs, and so a weakly coupled analysis is appropriate here. Along the flat directions the gauge symmetry is broken to $SU(5)$ while the  $SU(N-4)$ global symmetry is broken to $Sp(N-5)$. This is easily seen from the fact that
\begin{align}
A\bar{F}_i \bar{F}_j~=~\frac{1}{\sqrt{2}}\varphi^3 J_{ij},~~~~~~~~~i,j \leq N-5,
\end{align}
on the flat directions. There is also an unbroken $U(1)'$ global symmetry which is a linear combination of the original $U(1)$ and the $SU(N)$ global generator ${\rm diag}(5, \ldots ,5, -(N-4), \ldots , -(N-4))$. The dynamical scale of the unbroken $SU(5)$ is given by
\begin{align}
	\Lambda_{5}^{13} = \frac{\Lambda_{N}^{2N+3}}{({\rm Pf}' A \bar{F} \bar{F}) ({\rm Pf}' A)} \ ,
\end{align}
where the Pfaffians involve only the $N-5$ components, as indicated by the prime. This unbroken $SU(5)$ will have the same matter content as in the previous example, hence it confines and breaks supersymmetry at the scale $\Lambda_{5}$. Here and below, we absorb renormalization-scheme-dependent numerical constants \cite{Finnell:1995dr} into the definition of the scale $\Lambda$'s which does not affect any of the discussions below. The vacuum energy has a runaway dependence on $\varphi$ 
\begin{align}
	V \approx \Lambda_{5}^{4} 
	&= \left(\frac{\Lambda_{N}^{2N+3}}{\varphi^{2N-10}}\right)^{4/13}
	\label{eq:runaway}
\end{align}
without a stable ground state. It can be stabilized by adding $\lambda A \bar{F}_{i} \bar{F}_{j} J^{ij}$ to the superpotential \cite{Affleck:1983vc}.

With AMSB there is no need to add a tree-level superpotential, since the run-away behavior is stabilized by the scalar masses-squared along the flat direction
\begin{align}
	m_{A,\bar{F}_{i}}^{2} &= \frac{g^{4}}{(8\pi^{2})^{2}} 2 C_{i} (2N +3) m^{2}, 
\end{align}
where 
\begin{align}
C_{i} = \left\{ \begin{array}{lc} \frac{(N+1)(N-2)}{N} & \mbox{for } A, \\
\frac{N^{2}-1}{2N} & \mbox{for } \bar{F}_{i}.
\end{array} \right.
\end{align}
Note that Eq.~\eqref{eq:runaway} is a runaway {\it potential}\/ from dynamical supersymmetry breaking, not a {\it superpotential}\/. Therefore, there is no accompanying tree-level AMSB piece from \eqref{eq:AMSBW} to stabilize it. This is a key difference from the Affleck--Dine--Seiberg case of QCD-like theories \cite{Murayama:2021xfj}, and from the even-$N$ case discussed in the next section.
The theory has a stable ground state at
\begin{align}
	\varphi \approx \Lambda \left( \frac{4\pi\Lambda }{m} \right)^{13/(4N-7)} \gg \Lambda.
\end{align}
Therefore, the physics is weakly coupled at the scale $\varphi$ and our analysis is justified.

We now present a heuristic description of the dynamics in the Higgs picture, before passing to a purely gauge invariant formulation below. In this picture the UV theory has $A({\bf \frac{1}{2}N(N-1)})+\bar{F}({\bf \bar{N}})\times (N-4)$ chiral superfields. $N^2-1-24$ are eaten when the $SU(N)$ gauge symmetry is higgsed to $SU(5)$. Of the remaining ones, $15=A({\bf 10})+\bar{F}_{N-4}({\bf \bar{5}})$ are charged under the gauge $SU(5)$, and the other $\frac{1}{2}(N-4)(N-5)$ contain the Nambu-Goldstone bosons for the broken global symmetry
\footnote{The overall number of \textit{Nambu-Goldstone (NG) superfields} is given by $\frac{1}{2}\text{dim}(G^c/\tilde{H})$ \cite{LERCHE1984582}, where $G^c$ is the complexification of the original global symmetry $G$, and $\tilde{H}\supseteq H^c$ depends on the representation of the VEV. When $\tilde{H}=H^{c}$, each NG superfield contains one NGB. When $\tilde{H} \supset H^{c}$, there are fewer NG superfields, and hence some of them contain two NGBs to account for all the broken generators of $G/H$. In our case, we can do the counting in two steps. First we consider $SU(N-4)/SU(N-5)\times U(1)$ which gives the K\"ahler manifold ${\mathbb C}P^{N-5}$ and $N-5$ chiral superfields, each containing \textit{two} NGBs. In the second step we consider $SU(N-5)\times U(1)/Sp(N-5)$, which gives $\frac{1}{2}(N-5)(N-6)$ chiral superfields, each with a \textit{single} NGB and its superpartner. Overall, we get $\frac{1}{2}(N-5)(N-4)$, which is exactly the number of uneaten, $SU(5)$ neutral chiral superfields.}. 

The $SU(5)$ dynamics with $A({\bf 10})$ and $\bar{F}_{N-4}(\bf\bar{5})$ is along the lines of the previous section: $SU(5)$ is higgsed to $SU(4)$, supersymmetry is spontaneously broken, and there is a massless goldstino and an additional massless fermion. Anomaly matching is then satisfied by this fermion, together with the fermions in the first $N-5$ components of $\bar{F}_{N-4}$, which are neutral under the unbroken $SU(5)$ gauge group and remain massless. 
\begin{table}[t]
    \centerline{
	\begin{tabular}{|c|c|c|c||c|c|}\hline
	& $SU(N)$ & $U(1)$ & $SU(N-4)$ & $Sp(N-5)$ & $U(1)'$ \\ \hline
	$A$ & $\Yasymm$ & $-N+4$ & ${\bf 1}$ & ${\bf 1}$ & $-N+4$ \\ \hline
	$\bar{F}_{i}$ & $\overline{\Yfund}$ & $N-2$ & $\Yfund$ 
	& $\begin{array}{c}\Yfund\\ {\bf 1} \end{array}$ & 
	$\begin{array}{c}\frac{1}{2}(N-4) \\ \frac{(N+1)(N-4)}{2} \end{array}$ \\ \hline \hline
	$A \bar{F}_{i} \bar{F}_{N-4}$ & {\bf 1} & $N$ & $\Yfund$ 
	& $\begin{array}{c}\Yfund\\ {\bf 1}\end{array}$ & 
	$\begin{array}{c}\frac{N(N-4)}{2} \\ N(N-4) \end{array}$ \\ \hline
	\end{tabular}
	}
	\caption{Particle content of the $SU(N)$ theory with one anti-symmetric $A$ and  $N-4$ anti-fundamentals $\bar{F}_{i}$. $N=2n+1$ is odd. The operators $A\bar{F}_i \bar{F}_{N-4}$ are interpolating fields for massless fermions. Decompositions of fields under the unbroken $Sp(N-5) \times U(1)'$ symmetry are also shown. The $U(1)'$ charges are given separately for the fundamental and singlet of the unbroken $Sp(N-5)$ global symmetry.} \label{tab:SU(N)}
\end{table}

In gauge invariant language, the IR theory has $N-4$ massless fermions $A \bar{F}_i \bar{F}_{N-4}$ \footnote{Note that this notation is not a chiral superfield, but rather an interpolating field for the massless fermions in $\bar{F}_{N-4}$ written as gauge-invariant operators. This is consistent as supersymmetry is dynamically broken.} in the singlet and fundamental of the unbroken $Sp(N-5)$ global symmetry. These correspond to the $N-5+1$ massless fermions of the Higgs picture.

The massless fermions match all of the anomalies for the unbroken $Sp(N-5)\times U(1)'$:
\\\vspace{3pt}\\
$U(1)'\,{\rm gravity}^{2}:$
\begin{align}
	& {\bf \frac{1}{2}N(N-1)}(-N+4)+{\bf \bar{N}}(N-5)\frac{1}{2}(N-4)  \nonumber \\
	& + {\bf \bar{N}} \frac{1}{2}(N+1)(N-4) \nonumber \\
	&= (N-5)\frac{1}{2}N(N-4)+N(N-4),
\end{align}
$U(1)^{\prime 3}:$
\begin{align}
	&{\bf \frac{1}{2}N(N-1)}(-N+4)^{3} 
	+ {\bf \bar{N}}(N-5)\left( \frac{1}{2}(N-4)\right)^{3} \nonumber \\
	& + {\bf \bar{N}} \left( \frac{1}{2}(N+1)(N-4)\right)^{3} \nonumber \\
	&= (N-5)\left(\frac{1}{2}N(N-4)\right)^{3}+(N(N-4))^{3},
\end{align}
$U(1)'Sp(N-5)^{2}:$
\begin{align}
	{\bf \bar{N}} \frac{1}{2}(N-4) &= \frac{1}{2}N(N-4),
\end{align}
and $Sp(N-5)_{\rm Witten}:$
\begin{align}
	{\bf \bar{N}} &= 1~ (\mbox{mod 2}).
\end{align}

In addition, there are massless Nambu--Goldstone bosons of the $SU(N-4)/Sp(N-5)$ coset space, together with the Wess--Zumino--Witten term \cite{Wess:1971yu,Witten:1983tw} given that $\pi_5(SU(N-4)/Sp(N-5)) = \mathbb{Z}$ for the $SU(N-4)$ anomalies not contained in $Sp(N-5)$ induced by the one-loop diagrams of massive fermions \cite{DHoker:1984izu}. This is the second example of exact results on chiral gauge theories. 

When $m$ is increased, the field values approach the strong scale, and we lose control of the dynamics. Yet the non-trivial anomaly matching conditions depend only on the presence of the massless fermions and may well persist to the limit $m \gg \Lambda$. While there may be a phase transition that lifts the massless fermions discontinuously, our analysis provides a concrete suggestion for the dynamics of the non-supersymmetric chiral gauge theory, which should be checked explicitly by lattice methods. It would also be interesting to see whether the entire chiral compensator $\Phi$ could be embedded in a fully supersymmetric theory with spontaneous SUSY breaking giving rise to the scale $m$ governed by some holomorphic parameters of the UV complete theory. In that case one may perhaps argue against the presence of a phase transition in the $m\to \infty$ limit.

\section{Comparison to Tumbling}

While there is no controlled analysis for the  study of the dynamics of chiral non-supersymmetric gauge theories, there is a framework proposed by Raby,  Dimopoulos, and Susskind~\cite{Raby:1979my} that goes broadly under the name of ``tumbling". One first finds the Most Attractive Channel (MAC) among the charged fields and assumes that it condenses, breaking part of the gauge symmetry. This process is then iterated until one arrives at a QCD-like theory (or when the gauge group is fully broken). Applying this method of tumbling to the non-supersymmetric $SU(N)$ theory with chiral fermions $A\left(\Yasymm\right) +(N-4)\,\, \bar{F}\left(\overline{\Yfund}\right)$, one finds~\cite{Dimopoulos:1980hn} that the most attractive channel is the antisymmetric tensor $A^{ab}$ and the antifundamentals $\bar{F}_{bi}$ combined into fundamentals. This leads to a condensate
\begin{equation}
	\langle A^{ab} \bar{F}_{bi} \rangle \sim \Lambda^{3} \delta^{a}_{i} \neq 0, \quad~~~{i,a\leq N-4}\,.
	\label{eq:MAC}
\end{equation}
The above condensate breaks the $SU(N)$ gauge symmetry down to $SU(4)$ while leaving the (global) diagonal subgroup of $SU(N)\times SU(N-4)$ unbroken, locking the $N-4$ colors and flavors. The remaining $SU(4)$ symmetry is vectorlike and hence assumed confining. The global anomalies of the unbroken $SU(N-4)$ symmetry are matched by a massless fermion in the symmetric tensor representation of $SU(N-4)$ corresponding to the fermionic composite $A \bar{F}_{\{i,} \bar{F}_{j\}}$.  While this picture appears possible,  there are no controlled limits where the theory can be studied reliably along this line of analysis. 

Comparing to the results of our non-supersymmetric AMSB limit we can see that the above tumbling picture indeed appears to be incomplete. For the SUSY+AMSB case we have found that the remaining global symmetry is $Sp(N-5)\times U(1)'$ instead of the full $SU(N-4)\times U(1)$, with two massless fermion composites. It is actually easy to reconcile the tumbling picture with our SUSY inspired predictions. One needs to simply consider the second most attractive channel corresponding to two antifundamentals in the antisymmetric combination, and assume another condensate along this direction:
\begin{equation}
    \langle \bar{F}_{ai} \bar{F}_{bj} \rangle 
	\sim \Lambda^{3} J_{ab} J_{ij} \neq 0, \quad~~~{1\leq i,j,a,b\leq N-5}\,,
	\label{eq:condensate}
\end{equation}
in addition to the one in Eq.~\eqref{eq:MAC}. Since the gauge indices are anti-symmetric between $a$ and $b$ to make this channel attractive, the flavor indices $i$ and $j$ will also have to be anti-symmetric. Note that the symmetric combination $\delta_{ab} \delta_{ij}$ is repulsive and no condensate along that direction is expected. The condensate Eq.~\eqref{eq:condensate} breaks the global $SU(N-4)$ symmetry left by Eq.~\eqref{eq:MAC} further down to $Sp(N-5)$. 

The condensate in Eq.~\eqref{eq:condensate} is a good description in the weakly-coupled Higgs picture. When we increase $m$, the theory becomes strongly coupled and this description is no longer valid. Instead, we should refer to a gauge invariant order parameter for a consistent description, which in our case is
\begin{equation}
      \left< (\bar{F}_{i} \bar{F}_{j})_{[a,b]} (\bar{F}^{*k} \bar{F}^{*l})^{[a,b]} \right> \propto J_{ij} J^{kl}, ~{1\leq i,j,k,l\leq N-5}\,.
\end{equation}
 
The candidate Nambu--Goldstone bosons eaten by the $SU(N)/SU(4)$ massive gauge bosons are $A^{ab} \bar{F}_{bi}$ for the upper $(N-4)$-dimensional block and $A^{pb} \bar{F}_{bi}$ for the off-diagonal block where $p=1, \cdots, 4$ denotes the $SU(4)$ index. Finally $J^{ab}\bar{F}_{ai}\bar{F}_{bj}$ are the candidates for the uneaten Nambu--Goldstone bosons for the global $SU(N-4)/Sp(N-5)$ coset.

The condensate $A^{ab} \bar{F}_{bi}$ separates the $\bar{F}_i$ into the first $\bar{F}_{1\ldots(N-5)}$ components and $\bar{F}_{N-4}$. These end up in the $Sp(N-5)$ fundamental and singlet parts of the IR composite fermions $A \bar{F}_i \bar{F}_{N-4}$ - see Table~\ref{tab:SU(N)} for their charges.  Under the unbroken $SU(4)$ symmetry, the charged fermions decompose as $A({\bf 6}+{\bf 4})$ and $\bar{F}_{i}({\bf \bar{4}})$. These remaining degrees of freedom are vector-like, and become massive. The anomaly matching conditions remain the same and satisfied exactly as in the case of SUSY+AMSB. Therefore, this modified tumbling picture with the second condensate has the identical symmetry breaking pattern and massless fermion content as the SUSY theory with the AMSB, hence we find the two to be likely continuously connected. We expect this modified picture to provide the proper low-energy dynamics of the non-supersymmetric theory.

\section{even $N$}

\begin{table}[t]
    \centerline{
	\begin{tabular}{|c|c|c|c||c|c|}\hline
	& $SU(N)$ & $U(1)$ & $SU(N-4)$ & $Sp(N-4)$ \\ \hline
	$A$ & $\Yasymm$ & $-N+4$ & ${\bf 1}$ & ${\bf 1}$ \\ \hline
	$\bar{F}_{i}$ & $\overline{\Yfund}$ & $N-2$ & $\Yfund$ 
	& $\Yfund$ \\ \hline \hline
	$A \bar{F}_{i} \bar{F}_{j}$ & ${\bf 1}$ & $N$ & $\Yasymm$ 
	& $\Yasymm\oplus{\bf 1}$ \\ \hline
	Pf$A$ & ${\bf 1}$ & $-\frac{1}{2}N(N-4)$ & ${\bf 1}$ 
	& ${\bf 1}$ \\ \hline
	\end{tabular}
	}
	\caption{Particle content of the $SU(N)$ theory with one anti-symmetric $A$ and  $N-4$ anti-fundamentals $\bar{F}_{i}$. $N=2n$ is even. Decompositions under the unbroken $Sp(N-4)$ symmetry are also shown. }\label{tab:SU(even)}
\end{table}

Once again, we consider $SU(N)$ gauge theories with an anti-symmetric tensor $A$ and $(N-4)$ anti-fundamentals $\bar{F}_{i}$, but for even $N=2n$. See Table~\ref{tab:SU(even)} for quantum numbers. In this case, the $D$-flat directions break the gauge group to $Sp(4)=SO(5)$, whose gaugino condensate induces a dynamical superpotential \cite{Pouliot:1995me},
\begin{align}
	W = \left( \frac{\Lambda^{2N+3}}{({\rm Pf} A \bar{F} \bar{F}) ({\rm Pf} A)} \right)^{1/3} \ .
\end{align}
The AMSB Eq.~\eqref{eq:AMSBW} balances the superpotential against the supersymmetry breaking such that both $A\sim \bar{F} \sim \Lambda (\Lambda/m)^{3/2N}$ and all fermions acquire mass. The global $SU(N-4)$ and $U(1)$ symmetries are broken dynamically to $Sp(N-4)$ which does not have anomalies. The anomalies of broken symmetries are saturated by the Nambu--Goldstone bosons with the Wess--Zumino--Witten term given $\pi_5(SU(N-4)/Sp(N-4)) = \mathbb{Z}$. This is the third example of exact results on chiral gauge theories. 

Again this dynamics can persist to $m \gg \Lambda$. We can interpret the dynamics {in the non-SUSY limit} with the fermion bilinear condensates
\begin{align}
    \begin{array}{c}
    \langle A^{ab} \bar{F}_{bi}\rangle \sim \Lambda^3 \delta_i^a,~~~~\,\\[5pt]
    \langle \bar{F}_{ai} \bar{F}_{bj} \rangle \sim \Lambda^3 J_{ab} J_{ij}
    \end{array}
    \quad~~~{i,j,a,b\leq N-4}.
\end{align}
Note that in this case $\bar{F}_{N-4}$ is not singled out by the condensate, and there are no corresponding massless fermions. The remaining theory is again $SU(4)$ with $A({\bf 6}+{\bf 4})$ and $\bar{F}({\bf \bar{4}})$ which is vector-like and becomes massive. The massless degrees of freedom are chiral Lagrangian of $SU(N-4)\times U(1)/Sp(N-4)$ with the Wess--Zumino--Witten term.

\section{Conclusions}

In this paper, we outlined how dynamics of chiral gauge theories can be studied by perturbing the supersymmetric version with anomaly-mediated supersymmetry breaking. In particular, we worked out dynamics of $SU(N)$ gauge theories with an anti-symmetric tensor and $N-4$ anti-fundamentals. We came up with a consistent picture that connects supersymmetric gauge theories perturbed by the anomaly-mediated supersymmetry breaking to non-supersymmetric gauge theories. The symmetry breaking pattern suggested differs from that based on the original tumbling argument, which however can be extended to match the picture obtained here. It would be interesting to extend this analysis to other examples of chiral gauge theories. Ultimately, lattice gauge theory simulations will have the final verdict on the picture.

\section{Acknowledgments}

\begin{acknowledgments}

We thank Yuri Shirman for careful reading of the manuscript and useful comments. We also thank Jacob Leedom for pointing out typos. CC is supported in part by the NSF grant PHY-2014071 as well as the US-Israeli BSF grant 2016153. OT and HM were supported in part by the DOE under grant DE-AC02-05CH11231.
HM was also supported in part by the NSF grant
PHY-1915314, by the JSPS Grant-in-Aid for
Scientific Research JP20K03942, MEXT Grant-in-Aid for Transformative Research Areas (A)
JP20H05850, JP20A203, by WPI, MEXT, Japan, and Hamamatsu Photonics, K.K.
\end{acknowledgments}

\section{Appendix}\label{sec:Extraflavor}
\section{Calculable SUSY breaking minimum via extra flavor}

We have argued in the main text that SUSY $SU(5)$ with ${\bf 10+ \bar{5}}$, in addition to breaking SUSY spontaneously, also leaves the global $U(1)$ symmetry unbroken and produces a (non-supersymmetric) massless composite fermion that can be identified with $A\bar{F}\bar{F}$. In this appendix we confirm this statement by an explicitly study of the SUSY breaking vacuum. To do this, add an additional flavor to make the theory calculable $F+\bar{F}$ as first proposed in~\cite{Murayama:1995ng}. We will show that the theory with one added flavor breaks SUSY spontaneously, while preserving a global $U(1)$ with a massless fermion saturating its anomaly matching. Since we can decouple the extra flavor by taking its mass $m_M\rightarrow\infty$, the unbroken global $U(1)$ and the associated massless fermion persist in the theory with no extra flavor.  

We now present a detailed analysis of the theory with an extra flavor. The matter content and the global symmetries of this theory along with the 4 flat directions parametrizing the moduli space are given in Tab.~\ref{tab:extraflavor}. 

\begin{table}[t]
    \centerline{
	\begin{tabular}{|c||c|c|c|c|c||c|}\hline
	& $SU(5)$ & $SU(2)$ & $U(1)_M$ & $U(1)_Y$ & $U(1)_R$&$U(1)_{5}$ \\ \hline
	$A$ & $\Yasymm$ & $1$ & $2$ & $1$ & $0$ &$1$\\ \hline
	$\bar{F}_{i}$ & $\overline{\Yfund}$ & $\Yfund$ & $-1$ & $-3$ & $-6$ &$\begin{array}{c}2\\-3\end{array}$\\ \hline
    $F$ & $\Yfund$ & $1$ & $-4$ & $3$ & $8$ &$-2$\\
  \hline \hline
$B_1=AAF$ & $1$ & $1$ & $0$ & $5$ & $8$ &$0$\\ \hline
$H=A\bar{F}_1\bar{F}_2$ & $1$ & $1$ & $0$ & $-5$ & $-12$ &$0$\\ \hline
$M=F\bar{F}$ & $1$ & $\Yfund$ & $-5$ & $0$ & $2$&$\begin{array}{c}0\\-5\end{array}$\\ \hline
	\end{tabular}
	}
	\caption{Particle content of the $SU(5)$ theory with one extra flavor added, along with the quantum numbers of the flat directions. Note that we have chosen the most convenient basis of the $U(1)$ symmetries where $U(1)_M$ will be related to the unbroken $U(1)$ in the theory without the extra flavor. We also show the charges of the fields under the unbroken $U(1)_{5}$. These are given separately for the upper and lower components of the $SU(2)$ doublets.}\label{tab:extraflavor}
\end{table}
The flat directions corresponding to the gauge invariants in Tab.~\ref{tab:extraflavor} can be explicitly obtained by solving the $D$-flatness equations:
\begin{equation}
    A=\frac{1}{\sqrt{2}}\left( \begin{array}{ccccc} 0 & \alpha & 0 & 0 & 0\\ -\alpha & 0& 0 & 0 & 0 \\
   0 & 0 & 0 & \beta & 0 \\ 0 & 0  & -\beta & 0 & 0  \\ 0 & 0& 0& 0& 0 \\ \end{array} \right) \nonumber 
\end{equation}
\begin{equation}
    \bar{F}=\left( \begin{array}{cc} c_\delta~\omega  & s_\delta~\omega \\ -s_\delta~\gamma & c_\delta~\gamma \\
   0 & 0 \\ 0 & 0  \\ \delta_1 & \delta_2 \\ \end{array} \right), \ \ \  F = \left( \begin{array}{c} \mu \\ 0 \\ 0\\ 0 \\\frac{\beta\omega}{\gamma} \end{array} \right)  
\end{equation}
where $|\alpha|^2-|\gamma|^2= |\beta|^2$, $\mu = \pm \frac{\gamma}{\beta} \sqrt{\delta^2_1+\delta^2_2} $ and $\omega = \pm \frac{\gamma}{\beta}\sqrt{\beta^2+\delta^2_1+\delta^2_2}$, and also $c_\delta=\delta_1/\sqrt{\delta^2_1+\delta^2_2},\,s_\delta=\delta_2/\sqrt{\delta^2_1+\delta^2_2}$. 
To leading order, the directions $\delta_{1,2}$ coincide with the meson directions $M_{1,2}=F\bar{F}_{1,2}$. Going out along the $B_1$ and $H$ directions will break the $SU(5)$ gauge symmetry to $SU(2)$, which will produce a gaugino condensate. The resulting superpotential will be given by~\cite{Poppitz:1995fh} 
\begin{equation}
    W= \frac{\Lambda^6}{\sqrt{B_1 H}}
\end{equation}
leading to a runaway behavior for the $B_1$ and $H$ directions. These can be stabilized by adding the tree-level superpotential 
\begin{equation}
    W_{tree}=\lambda_1 B_1 + \lambda_2 H\,,
\end{equation}
where $B_1=AAF,\,H=A\bar{F}_1\bar{F}_2$. This results in the supersymmetric VEVs 
\begin{equation}
    B_1=\frac{\lambda_2^{\frac{1}{4}}}{\sqrt{2} \lambda_1^\frac{3}{4}} \Lambda^3,\ \ \
    H=\frac{\lambda_1^{\frac{1}{4}}}{\sqrt{2} \lambda_2^\frac{3}{4}} \Lambda^3
\end{equation}

If the dimensionless couplings $\lambda_{1,2}$ are $\ll 1$ then these VEVs will be larger than the dynamical scale $\Lambda$ leading to large masses of the broken gauge bosons, the corresponding eaten directions, while the radial directions will pick up a mass of order $\lambda^\frac{5}{6} \Lambda$. Mesons remain massless and saturate anomalies, \\\quad\\
$U(1)_M^{3}:$
\begin{align}
    {\bf 10} \times 2^3 + {\bf \bar{5}} \times 2 \times (-1)^3 + {\bf 5} \times (-4)^3
    = 2 \times (-5)^3
\end{align}
$U(1)_M {\rm gravity}^{2}:$
\begin{align}
    {\bf 10} \times 2 + {\bf \bar{5}} \times 2 \times (-1) + {\bf 5} \times (-4)
    = 2 \times (-5)
\end{align}\\
$U(1)_M SU(2)^{2}:$
\begin{align}
    {\bf \bar{5}} \times (-1)
    = (-5)
\end{align}
$SU(2)_{\rm Witten}:$
\begin{align}
    {\bf \bar{5}} = 1 ~ \mbox{(mod 2)}
\end{align}

Now we can add the final perturbation to the superpotential in the form of a mass to the extra flavor. Since there is only one fundamental $F$, without loss of generality we take it
\begin{equation}
    W_M = m_W M_1 = m_M F\bar{F}_1
\end{equation}
and assume that $m_M\ll \Lambda$. Supersymmetry is broken and the fermionic component of $M_1$ is the goldstino. 

The original paper \cite{Murayama:1995ng} did not work out the potential explicitly, nor mentioned the fermionic component of $M_2$ that stays massless. We will study this below.
For simplicity we will also assume that $m_M\ll \lambda^\frac{5}{6} \Lambda$ so that we can assume that in this perturbation the VEVs of $B_1, H$ (corresponding to the directions $\alpha,\beta,\gamma$) are held fixed. 
We can then find the explicit expression of the scalar potential to second order in $\delta_{1,2}$
\begin{equation}
    V= m^2_M \frac{\alpha^2}{\beta^2}\,\left(\beta^2+2\delta^2_1+\delta^2_2\right)\,.
\end{equation}
The potential is minimized at $\delta_1=\delta_2=0,\,V_{\text{min}}>0$, with positive masses. Hence, this is a calculable  stable SUSY breaking minimum. What happened to the $U(1)$ symmetries? The $U(1)_{Y,R}$ symmetries are explicitly broken by the tree level superpotential. $U(1)_M$ is broken explicitly by the addition of the $m_MM_1$ superpotential term, however a combination of the global $SU(2)$
and $U(1)_M$ is left unbroken, corresponding to 
\begin{equation}
    U(1)_{5}= 5 T_3 +\frac{1}{2} Q_M\ .
\end{equation}
Under this unbroken $U(1)_{5}$, the mesons $\delta_{1,2}$ have charges $0$ and $-5$, respectively. Because SUSY is spontaneously broken, the fermion component of $\delta_2$ remains massless even though its scalar component is massive, and it matches the 't Hooft anomalies for $U(1)_{5}$ as in Eq.~\eqref{eq:U(1)}.

This calculation establishes our basic claim of the presence of the unbroken $U(1)_{5}$ along with its corresponding massless fermion in the calculable regime. We can now take the limit $m_M\rightarrow\infty$ to decouple the extra flavor. While the position of the minimum is different when $m_M$ is no longer smaller than $\Lambda$, the unbroken $U(1)_{5}$ and its associated massless fermion persist in the decoupling limit by holomorphy. In the decoupling limit the massless fermion becomes $A\bar{F}\bar{F}$ [32], which has the same $U(1)_{5}$ charge $-5$ as $\delta_2$ and mixes with it due to dynamical SUSY breaking. 

\bibliographystyle{utcaps_mod}
\bibliography{chiralrefs}

\providecommand{\href}[2]{#2}\begingroup\raggedright\begin{thebibliography}{10}

\bibitem{Yang:1954ek}
C.-N. Yang and R.~L. Mills, ``{\em {Conservation of Isotopic Spin and Isotopic
  Gauge Invariance}},'' \href{http://dx.doi.org/10.1103/PhysRev.96.191}{Phys.
  Rev. {\normalfont \bfseries 96} (1954)  191--195}.

\bibitem{BCS}
J.~Bardeen, L.~N. Cooper, and J.~R. Schrieffer, ``{\em {Theory of
  Superconductivity}},''
  \href{http://dx.doi.org/10.1103/PhysRev.108.1175}{Phys. Rev. {\normalfont
  \bfseries 108} (1957)  1175}.

\bibitem{Nambu:1961tp}
Y.~Nambu and G.~Jona-Lasinio, ``{\em {Dynamical Model of Elementary Particles
  Based on an Analogy with Superconductivity. I}},''
  \href{http://dx.doi.org/10.1103/PhysRev.122.345}{Phys. Rev. {\normalfont
  \bfseries 122} (1961)  345--358}.

\bibitem{Nambu:1961fr}
Y.~Nambu and G.~Jona-Lasinio, ``{\em {Dynamical Model of Elementary Particles
  Based on an Analogy with Superconductivity. II}},''
  \href{http://dx.doi.org/10.1103/PhysRev.124.246}{Phys. Rev. {\normalfont
  \bfseries 124} (1961)  246--254}.

\bibitem{Ginsparg:1981bj}
P.~H. Ginsparg and K.~G. Wilson, ``{\em {A Remnant of Chiral Symmetry on the
  Lattice}},'' \href{http://dx.doi.org/10.1103/PhysRevD.25.2649}{Phys. Rev. D
  {\normalfont \bfseries 25} (1982)  2649}.

\bibitem{Kaplan:1992bt}
D.~B. Kaplan, ``{\em {A Method for simulating chiral fermions on the
  lattice}},'' \href{http://dx.doi.org/10.1016/0370-2693(92)91112-M}{Phys.
  Lett. B {\normalfont \bfseries 288} (1992)  342--347},
  \href{http://arxiv.org/abs/hep-lat/9206013}{{\normalfont \ttfamily
  arXiv:hep-lat/9206013}}.

\bibitem{Shamir:1993zy}
Y.~Shamir, ``{\em {Chiral fermions from lattice boundaries}},''
  \href{http://dx.doi.org/10.1016/0550-3213(93)90162-I}{Nucl. Phys. B
  {\normalfont \bfseries 406} (1993)  90--106},
  \href{http://arxiv.org/abs/hep-lat/9303005}{{\normalfont \ttfamily
  arXiv:hep-lat/9303005}}.

\bibitem{Narayanan:1994gw}
R.~Narayanan and H.~Neuberger, ``{\em {A Construction of lattice chiral gauge
  theories}},'' \href{http://dx.doi.org/10.1016/0550-3213(95)00111-5}{Nucl.
  Phys. B {\normalfont \bfseries 443} (1995)  305--385},
  \href{http://arxiv.org/abs/hep-th/9411108}{{\normalfont \ttfamily
  arXiv:hep-th/9411108}}.

\bibitem{Luscher:1998pqa}
M.~Luscher, ``{\em {Exact chiral symmetry on the lattice and the
  Ginsparg-Wilson relation}},''
  \href{http://dx.doi.org/10.1016/S0370-2693(98)00423-7}{Phys. Lett. B
  {\normalfont \bfseries 428} (1998)  342--345},
  \href{http://arxiv.org/abs/hep-lat/9802011}{{\normalfont \ttfamily
  arXiv:hep-lat/9802011}}.

\bibitem{Golterman:2000hr}
M.~Golterman, ``{\em {Lattice chiral gauge theories}},''
  \href{http://dx.doi.org/10.1016/S0920-5632(01)00953-7}{Nucl. Phys. B Proc.
  Suppl. {\normalfont \bfseries 94} (2001)  189--203},
  \href{http://arxiv.org/abs/hep-lat/0011027}{{\normalfont \ttfamily
  arXiv:hep-lat/0011027}}.

\bibitem{Golterman:2004qv}
M.~Golterman and Y.~Shamir, ``{\em {SU(N) chiral gauge theories on the
  lattice}},'' \href{http://dx.doi.org/10.1103/PhysRevD.70.094506}{Phys. Rev. D
  {\normalfont \bfseries 70} (2004)  094506},
  \href{http://arxiv.org/abs/hep-lat/0404011}{{\normalfont \ttfamily
  arXiv:hep-lat/0404011}}.

\bibitem{Grabowska:2015qpk}
D.~M. Grabowska and D.~B. Kaplan, ``{\em {Nonperturbative Regulator for Chiral
  Gauge Theories?}},''
  \href{http://dx.doi.org/10.1103/PhysRevLett.116.211602}{Phys. Rev. Lett.
  {\normalfont \bfseries 116} (2016) no.~21, 211602},
  \href{http://arxiv.org/abs/1511.03649}{{\normalfont \ttfamily
  arXiv:1511.03649}}.

\bibitem{Grabowska:2016bis}
D.~M. Grabowska and D.~B. Kaplan, ``{\em {Chiral solution to the
  Ginsparg-Wilson equation}},''
  \href{http://dx.doi.org/10.1103/PhysRevD.94.114504}{Phys. Rev. D {\normalfont
  \bfseries 94} (2016) no.~11, 114504},
  \href{http://arxiv.org/abs/1610.02151}{{\normalfont \ttfamily
  arXiv:1610.02151}}.

\bibitem{Raby:1979my}
S.~Raby, S.~Dimopoulos, and L.~Susskind, ``{\em {Tumbling Gauge Theories}},''
  \href{http://dx.doi.org/10.1016/0550-3213(80)90093-0}{Nucl. Phys. B
  {\normalfont \bfseries 169} (1980)  373--383}.

\bibitem{Dimopoulos:1980hn}
S.~Dimopoulos, S.~Raby, and L.~Susskind, ``{\em {Light Composite Fermions}},''
  \href{http://dx.doi.org/10.1016/0550-3213(80)90215-1}{Nucl. Phys. B
  {\normalfont \bfseries 173} (1980)  208--228}.

\bibitem{Bolognesi:2020mpe}
S.~Bolognesi, K.~Konishi, and A.~Luzio, ``{\em {Dynamics from symmetries in
  chiral $SU(N)$ gauge theories}},''
  \href{http://dx.doi.org/10.1007/JHEP09(2020)001}{JHEP {\normalfont \bfseries
  09} (2020)  001}, \href{http://arxiv.org/abs/2004.06639}{{\normalfont
  \ttfamily arXiv:2004.06639}}.

\bibitem{Bolognesi:2021yni}
S.~Bolognesi, K.~Konishi, and A.~Luzio, ``{\em {Probing the dynamics of chiral
  SU(N) gauge theories via generalized anomalies}},''
  \href{http://arxiv.org/abs/2101.02601}{{\normalfont \ttfamily
  arXiv:2101.02601}}.

\bibitem{Csaki:1997aw}
C.~Cs\'aki and H.~Murayama, ``{\em {Discrete anomaly matching}},''
  \href{http://dx.doi.org/10.1016/S0550-3213(97)00839-0}{Nucl. Phys. B
  {\normalfont \bfseries 515} (1998)  114--162},
  \href{http://arxiv.org/abs/hep-th/9710105}{{\normalfont \ttfamily
  arXiv:hep-th/9710105}}.

\bibitem{Gaiotto:2017yup}
D.~Gaiotto, A.~Kapustin, Z.~Komargodski, and N.~Seiberg, ``{\em {Theta, Time
  Reversal, and Temperature}},''
  \href{http://dx.doi.org/10.1007/JHEP05(2017)091}{JHEP {\normalfont \bfseries
  05} (2017)  091}, \href{http://arxiv.org/abs/1703.00501}{{\normalfont
  \ttfamily arXiv:1703.00501}}.

\bibitem{Tanizaki:2018wtg}
Y.~Tanizaki, ``{\em {Anomaly constraint on massless QCD and the role of
  Skyrmions in chiral symmetry breaking}},''
  \href{http://dx.doi.org/10.1007/JHEP08(2018)171}{JHEP {\normalfont \bfseries
  08} (2018)  171}, \href{http://arxiv.org/abs/1807.07666}{{\normalfont
  \ttfamily arXiv:1807.07666}}.

\bibitem{Bolognesi:2019fej}
S.~Bolognesi, K.~Konishi, and A.~Luzio, ``{\em {Gauging 1-form center
  symmetries in simple $SU(N)$ gauge theories}},''
  \href{http://dx.doi.org/10.1007/JHEP01(2020)048}{JHEP {\normalfont \bfseries
  01} (2020)  048}, \href{http://arxiv.org/abs/1909.06598}{{\normalfont
  \ttfamily arXiv:1909.06598}}.

\bibitem{Murayama:2021xfj}
H.~Murayama, ``{\em {Some Exact Results in QCD-like Theories}},''
  \href{http://arxiv.org/abs/2104.01179}{{\normalfont \ttfamily
  arXiv:2104.01179}}.

\bibitem{Randall:1998uk}
L.~Randall and R.~Sundrum, ``{\em {Out of this world supersymmetry
  breaking}},'' \href{http://dx.doi.org/10.1016/S0550-3213(99)00359-4}{Nucl.
  Phys. B {\normalfont \bfseries 557} (1999)  79--118},
  \href{http://arxiv.org/abs/hep-th/9810155}{{\normalfont \ttfamily
  arXiv:hep-th/9810155}}.

\bibitem{Giudice:1998xp}
G.~F. Giudice, M.~A. Luty, H.~Murayama, and R.~Rattazzi, ``{\em {Gaugino mass
  without singlets}},''
  \href{http://dx.doi.org/10.1088/1126-6708/1998/12/027}{JHEP {\normalfont
  \bfseries 12} (1998)  027},
  \href{http://arxiv.org/abs/hep-ph/9810442}{{\normalfont \ttfamily
  arXiv:hep-ph/9810442}}.

\bibitem{Pomarol:1999ie}
A.~Pomarol and R.~Rattazzi, ``{\em {Sparticle masses from the superconformal
  anomaly}},'' \href{http://dx.doi.org/10.1088/1126-6708/1999/05/013}{JHEP
  {\normalfont \bfseries 05} (1999)  013},
  \href{http://arxiv.org/abs/hep-ph/9903448}{{\normalfont \ttfamily
  arXiv:hep-ph/9903448}}.

\bibitem{Boyda:2001nh}
E.~Boyda, H.~Murayama, and A.~Pierce, ``{\em {DREDed anomaly mediation}},''
  \href{http://dx.doi.org/10.1103/PhysRevD.65.085028}{Phys. Rev. D {\normalfont
  \bfseries 65} (2002)  085028},
  \href{http://arxiv.org/abs/hep-ph/0107255}{{\normalfont \ttfamily
  arXiv:hep-ph/0107255}}.

\bibitem{Aharony:1995zh}
O.~Aharony, J.~Sonnenschein, M.~E. Peskin, and S.~Yankielowicz, ``{\em {Exotic
  nonsupersymmetric gauge dynamics from supersymmetric QCD}},''
  \href{http://dx.doi.org/10.1103/PhysRevD.52.6157}{Phys. Rev. D {\normalfont
  \bfseries 52} (1995)  6157--6174},
  \href{http://arxiv.org/abs/hep-th/9507013}{{\normalfont \ttfamily
  arXiv:hep-th/9507013}}.

\bibitem{Cheng:1998xg}
H.-C. Cheng and Y.~Shadmi, ``{\em {Duality in the presence of supersymmetry
  breaking}},'' \href{http://dx.doi.org/10.1016/S0550-3213(98)00539-2}{Nucl.
  Phys. B {\normalfont \bfseries 531} (1998)  125--150},
  \href{http://arxiv.org/abs/hep-th/9801146}{{\normalfont \ttfamily
  arXiv:hep-th/9801146}}.

\bibitem{Affleck:1983vc}
I.~Affleck, M.~Dine, and N.~Seiberg, ``{\em {Dynamical Supersymmetry Breaking
  in Chiral Theories}},''
  \href{http://dx.doi.org/10.1016/0370-2693(84)90227-2}{Phys. Lett. B
  {\normalfont \bfseries 137} (1984)  187}.

\bibitem{Leigh:1997sj}
R.~G. Leigh, L.~Randall, and R.~Rattazzi, ``{\em {Unity of supersymmetry
  breaking models}},''
  \href{http://dx.doi.org/10.1016/S0550-3213(97)00386-6}{Nucl. Phys. B
  {\normalfont \bfseries 501} (1997)  375--408},
  \href{http://arxiv.org/abs/hep-ph/9704246}{{\normalfont \ttfamily
  arXiv:hep-ph/9704246}}.

\bibitem{Murayama:1995ng}
H.~Murayama, ``{\em {Studying noncalculable models of dynamical supersymmetry
  breaking}},'' \href{http://dx.doi.org/10.1016/0370-2693(95)00744-6}{Phys.
  Lett. B {\normalfont \bfseries 355} (1995)  187--192},
  \href{http://arxiv.org/abs/hep-th/9505082}{{\normalfont \ttfamily
  arXiv:hep-th/9505082}}.

\bibitem{Poppitz:1995fh}
E.~Poppitz and S.~P. Trivedi, ``{\em {Some examples of chiral moduli spaces and
  dynamical supersymmetry breaking}},''
  \href{http://dx.doi.org/10.1016/0370-2693(95)01260-5}{Phys. Lett. B
  {\normalfont \bfseries 365} (1996)  125--131},
  \href{http://arxiv.org/abs/hep-th/9507169}{{\normalfont \ttfamily
  arXiv:hep-th/9507169}}.

\bibitem{Pouliot:1995me}
P.~Pouliot, ``{\em {Duality in SUSY SU(N) with an antisymmetric tensor}},''
  \href{http://dx.doi.org/10.1016/0370-2693(95)01427-6}{Phys. Lett. B
  {\normalfont \bfseries 367} (1996)  151--156},
  \href{http://arxiv.org/abs/hep-th/9510148}{{\normalfont \ttfamily
  arXiv:hep-th/9510148}}.

\bibitem{Georgi:1979md}
H.~Georgi, ``{\em {Towards a Grand Unified Theory of Flavor}},''
  \href{http://dx.doi.org/10.1016/0550-3213(79)90497-8}{Nucl. Phys. B
  {\normalfont \bfseries 156} (1979)  126--134}.

\bibitem{Elitzur:1975im}
S.~Elitzur, ``{\em {Impossibility of Spontaneously Breaking Local
  Symmetries}},'' \href{http://dx.doi.org/10.1103/PhysRevD.12.3978}{Phys. Rev.
  D {\normalfont \bfseries 12} (1975)  3978--3982}.

\bibitem{Finnell:1995dr}
D.~Finnell and P.~Pouliot, ``{\em {Instanton calculations versus exact results
  in four-dimensional SUSY gauge theories}},''
  \href{http://dx.doi.org/10.1016/0550-3213(95)00318-M}{Nucl. Phys. B
  {\normalfont \bfseries 453} (1995)  225--239},
  \href{http://arxiv.org/abs/hep-th/9503115}{{\normalfont \ttfamily
  arXiv:hep-th/9503115}}.

\bibitem{LERCHE1984582}
W.~Lerche, ``{\em On goldstone fields in supersymmetric theories},''
  \href{http://dx.doi.org/https://doi.org/10.1016/0550-3213(84)90336-5}{Nuclear
  Physics B {\normalfont \bfseries 238} (1984) no.~3, 582--600}.

\bibitem{Wess:1971yu}
J.~Wess and B.~Zumino, ``{\em {Consequences of anomalous Ward identities}},''
  \href{http://dx.doi.org/10.1016/0370-2693(71)90582-X}{Phys. Lett. B
  {\normalfont \bfseries 37} (1971)  95--97}.

\bibitem{Witten:1983tw}
E.~Witten, ``{\em {Global Aspects of Current Algebra}},''
  \href{http://dx.doi.org/10.1016/0550-3213(83)90063-9}{Nucl. Phys. B
  {\normalfont \bfseries 223} (1983)  422--432}.

\bibitem{DHoker:1984izu}
E.~D'Hoker and E.~Farhi, ``{\em {Decoupling a Fermion Whose Mass Is Generated
  by a Yukawa Coupling: The General Case}},''
  \href{http://dx.doi.org/10.1016/0550-3213(84)90586-8}{Nucl. Phys. B
  {\normalfont \bfseries 248} (1984)  59--76}.

\end{thebibliography}\endgroup

\end{document}